\documentstyle[aps,prb,psfig]{revtex}

\newcommand{\mathGraphic}[1]{\psfig{figure=#1}}

\def\cmm{{\rm cm}^{-1}}
\begin{document}

\title{Quantum tunneling dynamics using hydrodynamic trajectories. }
\author{Eric R. Bittner}
\address{Department of Chemistry, University of Houston, Houston, TX 77204}
\date{Submitted to  J. Chem. Phys.}
\maketitle
 \begin{abstract}
 In this paper we compute quantum trajectories arising from Bohm's 
 causal description of quantum mechanics.  Our computational 
 methodology is based upon a finite-element moving least-squares 
 method (MWLS) presented recently by Wyatt and co-workers (Lopreore 
 and Wyatt, Phys.  Rev.  Lett.  {\bf 82} , 5190 (1999)).  This method 
 treats the "particles" in the quantum Hamilton-Jacobi equation as 
 Lagrangian fluid elements which carry the phase, $S$, and density, 
 $\rho$, required to reconstruct the quantum wavefunctions.  Here, we 
 compare results obtained via the MWLS procedure to exact results 
 obtained either analytically or by numerical solution of the time 
 dependent Schr\"odinger equation.  Two systems are considered: 
 firstly, dynamics in a harmonic well and secondly tunneling dynamics 
 in a double well potential.  In the case of tunneling in the double 
 well potential, the quantum potential acts to lower the barrier 
 separating the right and left hand sides of the well permitting 
 trajectories to pass from one side to another.  However, as 
 probability density passes from one side to the other, the effective 
 barrier begins to rise and eventually will segregate trajectories in 
 one side from the other.  We note that the MWLS trajectories 
 exhibited long time stability in the purely harmonic cases.  However, 
 this stability was not evident in the barrier crossing dynamics.  
 Comparisons to exact trajectories obtained via wave packet 
 calculations indicate that the MWLS trajectories tend to 
 underestimate the effects of constructive and destructive 
 interference effects.
\end{abstract}

\section{Introduction}
Trajectory based constructions of quantum behavior are ubiquitous 
throughout quantum physics.  In the semi-classical limit, quantum 
dynamics is approximated by classical equations of motion whereby the 
transition amplitudes and wavefunctions are computed using the 
classical action connecting the initial and final states.  
Hydrodynamic constructions date back to early work by de 
Broglie\cite{ref01,ref02,ref03,ref04,ref05}and Madelung~\cite{ref5a} 
later by Bohm.  \cite{ref06,ref07} The so called quantum trajectories 
arising from this formalism have been the subject of a large number of 
ontological and philosophical papers seeking a causal interpretation 
of quantum mechanics.  While a comprehensive overview of these works 
is far beyond the scope of this paper, Holland's book provides perhaps 
the most comprehensive technical overview of the approach and contains 
many examples of how the formalism can be applied in a wide variety of 
cases.\cite{ref08} In short, the quantum trajectories themselves are 
relatively easy to compute once one has obtained a wavefunction 
solution of Schr\"odinger equation, $\psi$.  The velocity of a 
trajectory at a given point in space-time is computed as
\begin{eqnarray}
    v(x,t)=\frac{j(x,t)}{\rho (x,t)}
 \label{eq:0}
\end{eqnarray}
where $j(x,t)$ is the quantum current, and $\rho=|\psi|^2$.  In this 
"pilot wave" approach, $\psi$ acts as a guiding field for the 
trajectories.  This approach is useful in that once the wave function 
has been obtained, is generally easy to compute the trajectories.  
While there are a number of papers which have computed quantum 
trajectories having obtained the wave function, relatively little work 
has been done in developing computational methods based upon the 
description which does not rely upon first constructing the wave 
function.~\cite{ref09,ref10,ref11,ref12,ref13,ref14} However, Wyatt 
and co-workers have recently described a mesh-less finite-element 
method for integrating the de Broglie-Bohm equations using Lagrangian 
hydrodynamic trajectories.\cite{ref15,ref16,ref17} Similar methods are 
widely used in computational fluid dynamics (CFD) to simulate fluid 
flow dynamics in porous media (such as an oil reservoir) and other 
systems with complex topologies.~\cite{ref18} Wyatt's method 
represents the quantum density using a cloud of Lagrangian fluid 
points which themselves evolve according to the de Broglie-Bohm 
hydrodynamic equations of motion.  The method features a moving 
weighted least-squares (MWLS) approach to compute the various 
derivatives and gradients required by the de Broglie-Bohm equations.

In this work, we report on our implementation of the MWLS methodology 
and present an assessment of its difficulties and where improvements 
can be made.  Previous applications of the approach have focused 
entirely upon reactive scattering type calculations.  While this class 
of problems has its own set of associated difficulties, we have 
elected to focus upon the dynamics of systems trapped in various one 
dimensional potential wells.  These problems, while perhaps too 
idealistic, allow one to compare trajectories and results obtained via 
a "hydrodynamic" calculation to those obtained via analytical or 
numerical solution of the time-dependent Schrodinger equation.  
Specifically, we first examine the dynamics of harmonic oscillators 
and then move onto a more challenging problem of tunneling in double 
well potential.  Our results indicate that the methodology reproduces 
results obtained via more traditional grid based approaches; however, 
we note that the long time stability of the methodology needs to be 
improved before it can be used as an alternative to wave-packet based 
calculation.

\section{Quantum Equations of Motion}
The de Broglie-Bohm equations are derived directly from the time 
dependent Schr\"odinger equation.  The derivation is 
initialized by writing the wave function in polar form:
\begin{eqnarray}
  \psi (x,t) =R(x,t) e^{i S(x,t)/\hbar}.\label{eq:1}
\end{eqnarray}
When this substituted into the Schr\"odinger equation, the real and 
imaginary components can be collected into a continuity equation:
\begin{eqnarray}
    \partial_t \rho(x,t)+\frac{1}{m}\nabla \cdot( \rho\nabla S)  =
    \label{eq:2}
\end{eqnarray}
and the quantum Hamilton-Jacobi equation:
\begin{eqnarray}
   \partial_t S  + \frac{|\nabla S|^2}{2m} + V + Q = 0\label{eq:3}
\end{eqnarray}
where $Q$ is the so called quantum potential which is non-local and 
arises from the quantum kinetic energy operator in the Schr\"odinger 
equation.
\begin{eqnarray}
Q(x,t) = -\frac{{{\hbar }^2}}{2 m}\frac{1}{{\sqrt{\rho}}}{{\nabla }^2}{\sqrt{\rho}}\label{eq:4}
\end{eqnarray}
Taking the gradient of Eq.\ref{eq:3}, one obtains the equations of motion for the trajectories:
\begin{eqnarray}
{D_t}v(x,t)&=&-\frac{1}{m}\nabla (V(x(t)) + Q(x(t)))\label{eq:5}
\end{eqnarray}
where we define $x(t)$ and $v(x(t)) = \nabla S/m$ as the trajectory 
and velocity of Lagrangian fluid elements.  The notation $D_t f$ 
denotes the ``material'' or ``convective'' derivative of the function 
$f$ 
\begin{eqnarray}
    {D_t}f=\left(\partial_t+v\cdot \nabla \right)f
    \label{eq:5a}
\end{eqnarray}
which gives the rate of change of $f((x(t))$ as observed while 
moving with the particle along the trajectory, 
$x(t)$.\cite{ref20,ref21} Hence in the Lagrangian picture of fluid 
mechanics, we ``go with the flow".  The equation of continuity can also 
be written in terms of the convective derivative:
\begin{eqnarray}
{D_t}\rho  + (\nabla \cdot v) \rho  = 0\label{eq:6}
\end{eqnarray}
From this we can deduce a short time propagator for $\rho$:
\begin{eqnarray}
\rho (x,t+\delta t) = e^{-\nabla \cdot v\delta t} \rho(x,t)\label{eq:7}
\end{eqnarray}
which is local in space and is dependent upon the divergence of the 
velocity field at the point $x$.  To apply Eq.~\ref{eq:5} and 
Eq.~\ref{eq:7} in a computational scheme, we first discretize the 
system into an ensemble of Lagrangian fluid elements labeled by their 
position vectors, $x$.  These elements carry information regarding the 
local density about that point and the phase is determined by 
integrating Eq.~\ref{eq:4} along the path.  From these two bits of 
information, the quantum wavefunction can be reconstructed by writing 
\cite{ref17}
\begin{eqnarray}
\psi (x_i,t) ={\sqrt{\rho (x_i,0)}}
\exp\left(\frac{i}{\hbar}\int _{0}^{t}L({x_i},{v_i},\tau )d\tau
+\frac{i}{\hbar}S(x,0)\right)
\exp\left(-\frac{1}{2}\int_{0}^{t}\nabla \cdot v(\tau )d\tau\right)
\label{eq:9}
\end{eqnarray}
where 
\begin{eqnarray}
L(x,v,t)   =   \frac{1}{2}m   {v^2}   -   V   -   Q   \label{eq:10}
\end{eqnarray}
is the quantum Lagrangian and $S(x,0)$ represents any initial spatial 
dependent phase present in the initial wave function, such as $S(x,0) 
= \hbar kx$.

In order to compute expectation values we need to be able to integrate 
over the spatial domain spanned by either $\rho$ or $\psi$.  If we 
require that each trajectory element carry a volume element, $dx(t)$, 
then by normalization
\begin{eqnarray}
1=\int dx(0)\rho(x,0)   =   \sum_{i=1}^{N}\rho_i(0)dx_i(0)
=\sum_{i=1}^{N}\rho_i(t)dx_i(t),\label{eq:11}
\end{eqnarray}
where $\rho_i(t)=\rho(x_i(t))$ is the density carried by the $i$th trajectory.

To compute $dx_i(t)$ we need the Jacobian which transforms the volume 
element $dx_i(0)$ to the volume element, $dx_i(t)$ some time later 
along trajectory $x_i(t)$.
\begin{eqnarray}
J_i =\frac{\partial {x_i}(0)}{\partial {x_i}(t)}. \label{eq:12}
\end{eqnarray}
Taking the material time derivative of $J$ yields along each trajectory:
\begin{eqnarray}
D_t\log J_i   =   \nabla \cdot v_i\label{eq:13}
\end{eqnarray}
Thus, an amplitude element of the wavefunction is propagated along a 
trajectory $x_i(t)$ as
\begin{eqnarray}
dx_i(t)\psi_i(t) &=&
dx_i(0)
\sqrt{\rho_i(0)}
\exp\left(\frac{i}{\hbar}\int_{0}^{t}L({x_i},\dot{x}_i,\tau )d\tau +\frac{i}{\hbar}S(x_i,0)\right)  
\nonumber \\
&\times& \exp\left(-\int_{0}^{t}\nabla \cdot v_i(\tau)d\tau\right).
\label{eq:14}
\end{eqnarray}

A typical propagation cycle consists of computing the quantum 
potential from the density at each fluid element, taking the gradient 
of $Q$ and $V$ to compute the acceleration of the fluid element and 
computing divergence of the velocity field to determine how the 
density in given fluid element changes over one discrete time 
interval.  Thus, the trajectories, density, and velocity fields 
contain all of the essential information required to construct the 
full quantum mechanical wave function.

\section{Moving Least Squares Approximation}
Since our goal is to be able to compute the quantum trajectories 
without computing the wavefunction {\it a  priori}, we 
need to be able to be able to discretize the system into a set of 
Lagrangian fluid elements, compute the various derivatives required to 
compute the equations of motion and finally, advance the fluid 
elements to a new set of coordinates.  This we accomplish through the 
use of an adaptive meshless-cloud method described recently by Wyatt 
\cite{ref15,ref16,ref17}.  The method itself is described in 
detail by Liszka and co-workers \cite{ref18} and is 
similar to methods used extensively in computational fluid dynamics.  
First, we review the finite element approximation 
\cite{ref19} (FEA) and its implementation in a movable 
weighted least-squares scheme (MWLS).

In the moving least squares approximation, we assume that we have a 
function, $g= g(x)$ defined over a finite set of points 
and we seek an approximate value of $g$ at one of these 
points, $x_o$, based upon the values at neighboring 
points.  We assume that the function we seek is smooth enough to be 
expanded about the point $x_o$ in a finite 
polynomial basis
\begin{equation}
    p
    =  \left\{x,\frac{{x^2}}{2},\cdots 
\frac{{x^{n_{p}}}}{n_{p}!},\right\}\label{eq:15a}
\label{eq:polynomials}
\end{equation}
Thus, we can write
\begin{eqnarray}
f(x)= \sum_{i=1}^{n_p}a_ip_i(x-x_o)+f(x_o)  \label{eq:15}
\end{eqnarray}
where $a_i$ is a vector of coefficients.  For a simple polynomial as 
above, the derivatives of $f$
at $x_o$ are then just the coefficients 
themselves.  If we have a data point at $x_o$ and we 
want the approximation to pass though the remaining data points, we 
can write a set of least-squares equations
\begin{eqnarray}
f_j=\sum_{i=1}^{{n_p}}{a_i}{p_i}({x_j}-{x_o}),
    \label{eq:16}
\end{eqnarray}
where $f_j= f(x_j)-f(x_o)$, $p_i(x_j-x_o)$ is the $i$th polynomial 
basis member evaluated at ${x_j}-{x_o}$, and $a_i$ are the expansion 
coefficients.  In other words, we seek a vector $a$ which 
satisfies the linear equation:
\begin{eqnarray}
f=P\cdot a
\end{eqnarray}
where $f$, $P$, and $a$ are the terms in Eq.~\ref{eq:16} written in matrix/vector form.

In order to solve for $a$, we need at least as many data points in the 
neighborhood about $x_o$ as we have terms in the 
polynomial expansion.  Rather than trying to pick out exactly the 
right number of points, the most efficient procedure is to simply 
select more data points than basis functions.  Since the system of 
equations is overdetermined, $P$, is a rectangular matrix and 
we use must either singular value decomposition or other 
pseudo-inversion method to invert $P$.  Furthmore, we can 
improve the stability of the least squares procedure by weighting each 
data point so that points farther away from the central point receive 
the least weight.  Thus, for the "weighted" least squares procedure we 
solve
\begin{eqnarray}
f^{\Omega }  =   P^{\Omega }\cdot a
    \label{eq:wls}
\end{eqnarray}
where $f_j^\Omega=(f_j-f(x_o))\omega_j$, 
$P_{ij}^\Omega=\omega_jP_{i j}$, $a$
is a vector of undetermined coefficients, and $\omega_j$
is the weight assigned to point $x_j$.   

We have found that the convergence of the least-squares procedure is 
greatly improved by using a logarithmic form of the 
density\cite{ref14,ref16,numrec},
\begin{eqnarray}
\rho =e^{g(x)},
    \label{eq:exprho}
\end{eqnarray}
where $g$ is a polynomial of order $n_p$.  
This representation is a useful way to transform a non-linear model to a 
linear one.  For example, in computing the quantum potential we can write
\begin{eqnarray}
\frac{1}{\sqrt{\rho}}
\nabla^2\sqrt{\rho } = e^{-g/2}\nabla^2e^{g/2}
  \label{eq:qpot1}
\end{eqnarray}
Working through the derivative terms yields:
\begin{eqnarray}
e^{-g/2}\nabla^2e^{g/2} = \frac{1}{2}\left(g_{,\mu \mu }+\frac{1}{2}g_{,\mu}^{2}\right)   
    \label{eq:qpot2}
\end{eqnarray}
where the "comma" delimited subscripts in $g_{,u}=\partial_\mu g $ denote taking derivative
with respect to coordinate $\mu$.

Finally we write the equations of motion for the velocity in terms of 
the derivatives of the various fields in component form:
\begin{eqnarray}
    m a_{\mu }(x)   &=&   -\partial_\mu \left(V(x)- \frac{\hbar^2}{4  m}
    \left(g_{,\nu \nu} +\frac{1}{2}g_{,\nu }^{2}\right)\right) \\
\rho (x,t + \delta t)&=& e^{-\partial_\mu v_\mu(x)\delta t}\rho (x,t).
\end{eqnarray}

Some final notes regarding our implementation of the MWLS scheme are 
also in order.  We propagate trajectories using the sympletic Verlet 
algorithm ~\cite{ref22}.  Convergence with respect to time step is checked by 
comparing the results of two 1/2 time steps to one full time step.  If 
the differences between position, energy, velocity, or $\rho$
computed via the two separate procedures varied by more than 
$1:10^6$, the shorter time step results were taken, 
$\delta t$ was reduced by a factor of 0.75 and propagation 
continued.  However, if the longer time step results were sufficiently 
accurate, $\delta t$ was increased by a factor of 2 for the next 
iteration cycle.  For the least squares fitting procedures, we found 
that a Hermite polynomial basis provided the most robust basis as 
measured by the $\chi^2$ for each fit.  We also 
weighted the fits using a Gaussian weighting function centered about 
$x_o$ adjusted so that the point farthest from 
$x_o$ received a weight of 0.01.  For the one 
dimensional results presented here, our code reproduces results 
obtained via Wyatt's MWLS particle code described in 
Ref.\cite{ref15}

\section{Example Calculations}
\subsection{Harmonic Oscillator}
As a primary example, we consider the quantum trajectories for a 
harmonic oscillator and compare these to the classical trajectories 
starting from identical starting points.  In Fig.  1 we show the time 
evolution of an ensemble of quantum trajectories for a harmonic system 
with period of $\tau= 2\pi/\omega= 888.57 a.u.$ 
Unless otherwise noted explicitly, we shall use atomic units from here 
on, setting $\hbar = 1$.  The first two cases (A \& B) are 
the quantum trajectories for a particle with mass $m=2000 a.u$
and with mass $m=200 a.u.$ respectively, 
each starting with identical initial quantum densities 
\begin{eqnarray}
\rho(x) =e^{-\beta (x-{x_o})^2}    
\label{eq:rhoinit}
\end{eqnarray}
with $x_o=3.0$ and $\beta=0.3$.

In each of the cases shown here, we used 100 points spaced evenly 
about the center of the initial density.  In fourth plot (D) we show 
the trajectories obtain from a purely classical calculation involving 
the same 100 points.  The effect of the quantum potential is clearly 
seen as the trajectories traverse the minimum of the well at $x=0$. 
For the heavier mass (in A), the trajectories tend to focus 
at the well minimum, just as in the classical calculation (in D).  
However, the trajectories {\em do not cross} each other as do the 
classical trajectories.  This is a crucial characteristic of Bohmian 
trajectories.  In essence, the density must be single valued and the 
Jacobian of the volume element carried by each point must remain 
positive definite for all times.  Hence, if any pair of trajectories 
were to cross, these conditions would be violated.  For case B, in 
which $m=200 a.u$ , the quantum potential is $10\times$ stronger 
than in case A and hence the trajectories become more 
diffuse as the particle traverses the bottom of the well.  
Interestingly enough, if we instead choose $\beta = 1/m\omega $, 
the trajectory lines remain evenly 
spaced throughout the calculation as shown in C.

This last result can be understood from the fact that the wavepacket 
used in Fig.  1.C is a coherent state and hence 
the time evolution of the density is given by
\begin{eqnarray}
\rho (x,t) = e^{-m\omega (x-{x_o}\cos(\omega t))^2}
    \label{eq:coherent}
\end{eqnarray}
where $x_o$ is the original centroid of the wavepacket.  The quantum 
potential can be easily derived from Eq.  26 and the action along any 
given Bohmian trajectory is given by~\cite{ref08}
\begin{eqnarray}
S(x,t)=\frac{1}{2}\hbar \omega  t
-\frac{1}{2}m   \omega    \left(2   x   {x_o}\sin(\omega
   t)-\frac{1}{2}x_{o}^{2}\sin(2   \omega    t)\right).
   \label{eq:27}
\end{eqnarray}

Thus, each trajectory is a solution of
\begin{eqnarray}
m\dot{x} = \partial_x S   = -m   \omega    {x_o}\sin\omega    t. \label{eq28}
\end{eqnarray}
and belongs to the family of cosine curves of period 
$\omega$:
\begin{eqnarray}
x_i(t) = x_i(0) +x_o(\cos(\omega t) -1) .
    \label{eq:29}
\end{eqnarray}
{\em I.e.}  the Bohmian particle undergoes simple 
harmonic motion of amplitude $x_o$ about 
the point $x_i(0)-x_o$.
Furthermore, the energy along a given Bohmian trajectory is computed 
by taking the time derivative of the action (Eq.~\ref{eq:27}):
\begin{eqnarray}
E(x,t) =  \partial_t S  = \frac{\hbar \omega}{2}
-\frac{1}{2}  m \omega^2
(2 x x_o\cos(t   \omega ) -x_o^2\cos(2   t   \omega )   ).
    \label{eq:30}
\end{eqnarray}
Note that this is not constant in time as would be expected for a 
purely classical system.  The quantum force thus acts as a time 
dependent driving force on the ensemble of trajectories.  If we were 
to have chosen the initial center of the density to be at exactly $x 
=0$, then $E =\hbar\omega/2$ and the quantum force would exactly 
counter-balance the classical potential force.  In essence, the 
particles would remain fixed in their spatial positions.

In the case of the quantum harmonic oscillator, the quantum 
trajectories computed using the MWLS procedure agree with the exact 
analytical results one can obtain for this system.  Furthermore, the 
procedure appears to be stable for long periods of time.  We have 
carred out calculations up to 10 oscillation periods without 
considerable accumulation of error.  We now move on to the more 
formidable problem of tunneling in a double well potential.

\subsection{Quantum Tunneling in a Double Well Potential}
Tunneling and barrier penetration represent phenomena which are unique 
to quantum mechanics.  It is well understood from classical mechanics 
that if a particle lacks sufficient energy to surmount a potential 
barrier, the particle will be reflected and the barrier is 
impenetrable.  However in quantum mechanics, a particle can "pass 
through" a barrier.  This can be rationalized in a variety of ways.  
Perhaps the most consistent is that if the particle is in brought to a 
halt at the classical turning point, both its position and momentum 
are defined to exact precision, which would violate the uncertainty 
principle.  In the quantum trajectory approach, the quantum potential acts 
to enforce this behavior by effectively "lowering" the barrier to 
permit trajectories to penetrate through the barrier and raises the 
barrier once sufficient amplitude has passed.  To see how this comes 
about in the context of a particle based approach, let us consider 
tunneling through the double well potential:
\begin{eqnarray}
V(x)   =   a   {x^4}-   b {x^2} \label{eq:31}
\end{eqnarray}
In this example, we take $ a =0.007$ and $b= 0.01$ in atomic units, 
which gives two tunneling states below the barrier at $E= 
-369.827\cmm$ and -313.918 $\cmm$ and a barrier height of 
$V_b=786.24\cmm$.  The initial Gaussian density was centered about the 
center of the right side well at $x_o=\sqrt{b/2a}$ with $\beta=\sqrt{4 
b m}$ corresponding to a harmonic approximation to a state localized 
in the right hand well.  In the lower frame of Fig.  2 we show quantum 
trajectories for the initial crossing of the wavefunction from the 
right to the left side of the well and in Fig.  3 we show snapshots of 
the quantum density, the quantum potential, $V+Q$, and $V(x)$ for 
various times.  The trajectories cross from $+x$ to $-x$ corresponding 
to top to bottom in Fig 2.  The initial expansion of the paths is 
primarily due to the expansion of the wavefunction in the well.  Since 
the initial state is not in a stationary state of the potential, there 
is a net force on the particles due to the difference between the $Q'$ 
and $V'$.  Once the particles have passed through the barrier region 
(at $x=0$), they encounter the repulsive potential wall and are 
reflected.  Note that the particles penetrate fairly deeply into the 
repulsive region.  However, they carry very little net density.  
Unfortunately as far as our calculations are concerned, we run into 
numerical difficulties in computing the quantum potential in this region.  
As a result, the congruency condition fails to hold and trajectories 
begin to cross (after $t = 650$) and our results are meaningless 
beyond this point.

However, prior to this point in time, we can compute the effective 
barrier height due to the quantum potential at $x=0$.  In Fig.  4 we plot 
the effective barrier height, $V_{eff}=Q(0)+{V_b}$, as a function of 
time.  As indicated above, the quantum potential from the initial Gaussian 
distribution effectively "lowers" the barrier allowing trajectories to 
pass from the right to the left.  However, once the initial expansion 
of the density is accomplished, the barrier begins to rise and the 
effective barrier height appears to approach an asymptotic value.

If we assume that the lowest two eigenstates of the double well can be 
approximated as symmetric and antisymmetric linear combinations of 
ground state harmonic oscillators centered about the left and right 
minima in the well, we can compute the approximate quantum potential at 
$x=0$ quite easily.  Let us write the approximate wavefunction as 
(neglecting a common phase factor),
\begin{eqnarray}
\psi(x,t)   =  
\frac{1}{\sqrt{2}}
(e^{-i \omega t}\phi_{+}(x)  +
   e^{+i\omega t}\phi_{-}(x)),
    \label{eq:34}
\end{eqnarray}
where $\phi_\pm(x)$ are the symmetric and anti-symmetric tunneling states 
split by $\hbar\omega =(E_-E_+)/2$.  Taking,
\begin{eqnarray}
\phi_\pm(x) \approx\frac{1}{\sqrt{2}}
(\phi_R(x) \pm \phi_L(x)),
    \label{eq:35}
\end{eqnarray}
and using the approximation
\begin{eqnarray}
\phi_{R,L}(x)  = 
\left(
\frac{m \omega}{\pi \hbar }
\right)^{1/4}
e^{-m \omega_o(x\pm{x_o})^2/2\hbar },   
    \label{eq:36}
\end{eqnarray}
the quantum potential at the barrier is given by
\begin{eqnarray}
Q(0)  =
\left(\frac{\hbar \omega_o}{2} +   \frac{1}{4}m  \omega_o^{2}
x_o^2( \cos(4  t   \omega )-3)\right)
    \label{eq:37}
\end{eqnarray}

Thus, the effective barrier separating the right and left sides of the 
system is at its minimum when $\rho $ is localized in the right or 
left hand well and greatest when $\rho $delocalized in a 50:50 mixture 
of right and left hand states.  Moreover, as the separation between 
the respective wells increases, the harmonic oscillator approximation 
to the lowest eigenstates becomes better and better.  Correspondingly, 
the quantum potential becomes the parabola,
\begin{eqnarray}
Q(x) \approx    \frac{\hbar \omega_o}{2}
-  \frac{1}{2}m  
\omega_o^2(x\pm x_o)^2.
    \label{eq:38}
\end{eqnarray}

Since the quantum force and the classical potential force will be 
nearly equal and opposite, the particles will remain almost 
motionless in the initial well.

\section{ Pilot wave trajectories}
As noted above, the trajectories themselves can be obtained by 
propagating a solution to the time-dependent Schr\"odinger equation.  
In this ``pilot wave'' scheme, the quantum potential and equations of 
motion for the Bohmian particles may be computed directly from the 
wave function itself.  Alternatively, if we have an accurate 
representation of $\psi$ and can compute $v[\psi]$ via Eq.~\ref{eq:0}, 
we can obtain what should be a set of ``exact'' Bohmian trajectories 
associated with the quantum wavefunction.  To compute such 
trajectories for the double well example shown above, we used a 
discrete variable representation \cite{ref23} (DVR) of 
Gauss-Tchebychev quadrature points to represent both the wave packet 
and the time-evolution operator on a finite spatial grid of 200 
quadrature points spanning 2.5 Bohr on either side of the barrier.  
For dynamics at low total energy, this representation is certainly 
more than adequate as we were able to converge the lowest 20 
eigenstates of this well to $1:10^6$.  Because the instantaneous 
position of a Bohmian particle at $x(t)$ will not generally correspond 
to the position of a quadrature point, the velocity can not be 
computed directly from the DVR wavefunction, which is only defined on 
the quadrature points them selves.  Rather, we transform $\psi$ from 
the DVR to a finite basis representation (FBR) using the unitary 
transformation,
\begin{eqnarray}
\psi^{FBR}   =   U\cdot\psi^{DVR},    
    \label{eq:39}
\end{eqnarray}
where $U$ is the unitary transformation associated with a $N$-point Gauss-Tchebychev DVR.
\begin{eqnarray}
U_{i j}   =  \sqrt{\frac{2}{N+1}}\sin\left( i  j  \frac{\pi }{N+1}\right).    
    \label{eq:40}
\end{eqnarray}
The elements of the vector $\psi^{FBR}$ are the expansion coefficients of the 
wavefunction in a finite polynomial basis.
\begin{eqnarray}
    \psi(x,t) = \sum_{i=1}^{n}\psi_i^{FBR}(t) {T_i}(x).
    \label{eq:41}
\end{eqnarray}

For the Gauss-Tchebychev quadrature scheme used here, basis functions are:
\begin{eqnarray}
T_i(x) = \sqrt{\frac{2}{\Delta }}
\sin(i \pi (x-\Delta /2)/\Delta ).
    \label{eq:42}
\end{eqnarray}

Eq.  37 gives the wavefunction at any point $x \in 
[-\Delta/2,\Delta/2]$ and we can use this information to compute the 
velocity of a particle guided by $\psi$.  In the top frame of Fig.  2 
we show the Bohm trajectories for the double well system choosing 
identical conditions as above, this time computed via the DVR pilot 
wave based procedure.  For clarity we show only those trajectories for 
particles starting to the left of the initial centroid of the density.  
At long times, the trajectories transmitted to the left hand well tend 
to bunch together.  This is due to the fact that the finite basis 
functions in Eq.  38 go to zero at $x=\pm\Delta/2$ and force the 
system to have an artificial node at the boundary points.  
Consequently, the quantum potential is infinitely repulsive at the end 
of the grid and all particle trajectories are repelled from these 
regions.  For practical purposes, the density carried by these 
trajectories is negligible and the overall effect on the wavepacket 
dynamics is minimal.

Notice, however, that between $t = 600$ and $t = 800$, a number of 
trajectories near the barrier at $x=0$ are deflected from each other.  
This is due to the fact that $\rho$ develops oscillatory structure due 
to constructive and destructive interferences between on coming and 
reflected components of the wavefunction near the barrier.  The 
density eventually develops a node at about $t = 900$ a.u. 
Surprisingly, and disturbingly, this behavior is absent in the MWLS 
trajectories.  In Figure 5 and Figure 6 we compare the evolution of 
$\rho$ using the DVR $(\rho^{DVR})$ and MWLS $(\rho^{MWLS})$ methods.  
Initially, the agreement is quite good.  At longer times the agreement 
is very poor with $\rho^{MWLS}$ even "looping back" through itself.  
At intermediate times, as $\rho$ is reflected by the repulsive barrier 
in the $-x$ half of the double well, oscillations develop in 
$\rho^{DVR}$ whereas no such oscillations are seen in $\rho^{MWLS}$.  
Since the quantum potential is a measure of the local curvature of 
$\rho$, these oscillations translate into attractive and repulsive 
regions in the quantum potential.  Eventually as $\rho^{DVR}$ goes on 
to form nodes (at approximately $ t = 700$ au and 900 au trajectories 
are guided away from such regions via the quantum force.  This 
accounts for the significant deflections in the DVR trajectories but 
does not tell us why this fails to occur in the MWLS case.

The answer is basically that of sampling resolution.  As the MWLS 
trajectories evolve, they by in large tend to spread apart.  This 
restricts our ability to resolve any structure in $\rho^{MWLS}$ which 
occurs on lengthscales finer than the distance of separation of the 
MWLS trajectories at a given point in time.  Hence, the oscillatory 
structure seen in the DVR calculations (e.g. at $t$=550 au) are too 
fine to be resolved by the MWLS trajectories.  Thus, the quantum 
potential will be too smooth in this region.  The DVR calculations are 
also band width limited due to the finite spacing of the DVR points.  
However, in the DVR grid used in these examples, this spacing is fixed 
at $\delta x =0.049$ Bohr, where as in the MWLS case the spacing 
between trajectories in the region of the barrier at $t=550$ is 
$\approx 0.10$ Bohr, which slightly courser than the structure seen in 
$\rho^{DVR}$ in this region.  Consequently, the MWLS trajectories lose 
band width resolution precisely where it is needed the most in this 
case.

Let us compare the DVR and MWLS based trajectories on a head to head 
basis.  Four sets of trajectories extracted from Fig.  2 are shown in 
Figure 7.  First, for trajectories originating near the leading edge 
of the MWLS grid, we expect that the two results should deviate rather 
early due to the fact that the fitting procedure used to compute the 
quantum potential using the MWLS is least accurate near the end of the 
grid (Trajectory \#{} 9).  Surprisingly, however, the agreement is 
quite good.  In fact, it is appearent that the deviatiation is more 
likely due to the artificial boundary conditions imposed by the basis 
function used to represent the quantum wave packet in the DVR 
calculations.  For trajectories originating near the maximum of the 
initial wavepacket (Trajectory \#{}50), we see quantative agreement 
between DVR and MWLS based trajectories since $\rho $ is smooth 
function of $x$ in this region and remains so over the course of both 
calculations.  However, let us compare the two trajectories labeled 
\#{}38 and \#{}39 from each calculation.  These two trajectories 
originate side by side separated by $\delta x = 0.02 Bohr$.  In the 
DVR calculation, these trajectories are deflected by the oscillations 
in $\rho$ with \#{}38 deflected slightly towards the $-x$ direction 
and \#{}39 in the $+x$ direction.  These deflections are not observed 
in the corresponding MWLS trajectories.

\section{Discussion}

In this paper, we present our implementation and computational 
analysis of the de Broglie-Bohm hydrodynamic equations using a 
particle based description.  The formalism itself allows an elegant 
interpretation of quantum dynamics in geometric terms.  Instead of 
wave functions, we have geometric ray lines.  Surprisingly, this 
intuitive connection between quantum wave mechanics and geometry has 
not made it into the ``main stream'' even though the original seeds 
for this interpretation can be found at the very advent of quantum 
theory.  However, we do note that this view point is seeing a 
resurgence in popularity as measured by the number of papers which 
have used the Bohmian construction in vastly different areas of 
quantum physics.

What is important, however, to point out that we are only showing the 
best case scenarios in this paper.  We have found that trajectories 
tend to cross cases in which $\rho$ is very small, trajectory lines 
are far apart, or when the potential is significantly anharmonic.  
Some of this is exhibited in Figures 2 and 3.  This is most certainly a 
result of numerical instabilities in the MWLS procedure which we have 
elaborated upon in this paper.

We also run into particular dificulties in dealing with nodal points in the 
density, in other words, when $\rho(x)=\psi(x)|^2 = 0$.  At such 
points, it is not possible to fit the coefficients of a polynomial 
expansion using points on either side of the node since the function 
does not have continuous derivatives at the node.  We have tried 
various computational alternatives to dealing with such points.  One 
alternative is to use only trajectories and densities on one side of the 
node or the other when computing $Q(x)$.  Under this approach we 
compute derivatives by approaching the node from the right and from 
the left.  Another alternative we have tried is to introduce a 
$\log(x-{x_n})$ basis function for stars which include nodal points.  
Both of these approaches give very good results for cases in which the 
initial wavefunction was taken as an odd parity eigenstate in a 
symmetric well.  However, both cases require prior knowledge of the 
location of the node and that parity be a constant of the motion so 
that the node does not move.  For cases in which parity is not a 
constant of the motion, nodal points can occur over the natural course 
of the evolution of the quantum wavefunction, even if the initial 
wavefunction has no nodes initially.  We speculate that a solution may 
be to construct
\begin{eqnarray}
    R_i = \sqrt{\rho_i}
 e^{i \mu \pi }
    \label{eq:45}  
    \end{eqnarray}
along each $x_i(t)$ trajectory where $\mu = 0,1,2,\ldots$ accounts for 
the branch cut taken when evaluating $\sqrt{\rho}$ on either side of a 
node.  ~\cite{ref24} In essence, on one side of a node $R > 0$ and 
$\mu =0,2,...$, while on the other side, $R< 0$ and $\mu =1,3,\ldots$ 
This would preserve analyticity across the node allowing us to 
calculate $Q$ using a simple polynomial basis.  At present, the 
treatment of nodes and their evolution is an open problem which we 
will take up in a future work.\cite{ref25}

\acknowledgments This work was supported by grants from the Robert A. 
Welch Foundation and the National Science Foundation.  The author also 
wishes to thank Prof.  Bob Wyatt (Univ.  of Texas), Prof.  Don Kouri 
(U. Houston) and Dr.  Pablo Yepes (Rice) for stimulating discussions.

\begin{figure}
 \caption{Quantum trajectories for harmonic oscillators.In each 
 case,the oscillation period,$\tau =2 \pi/\omega =888.57 au$. 
 In cases A \&{} D ,$m=2000 au$ while in case B, 
 $m =200 au$.  Case D is a set of classical trajectories ($Q=0$) 
 for this system.  }
\mathGraphic{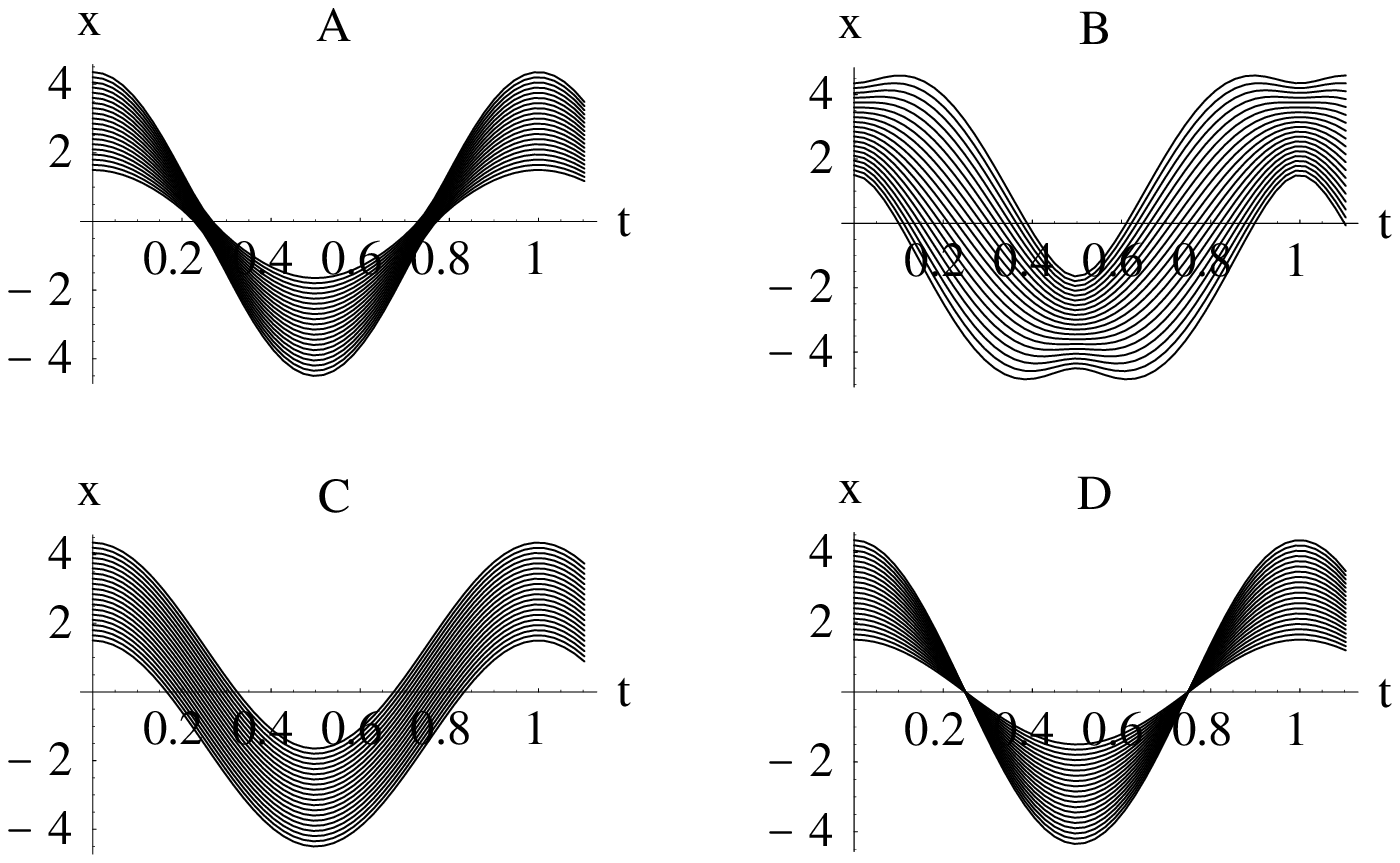}
\end{figure}

\begin{figure}
\caption{ Bohmian trajectories computed via wavepacket propagation 
(DVR) versus particle based trajectory method (MWLS).}
\mathGraphic{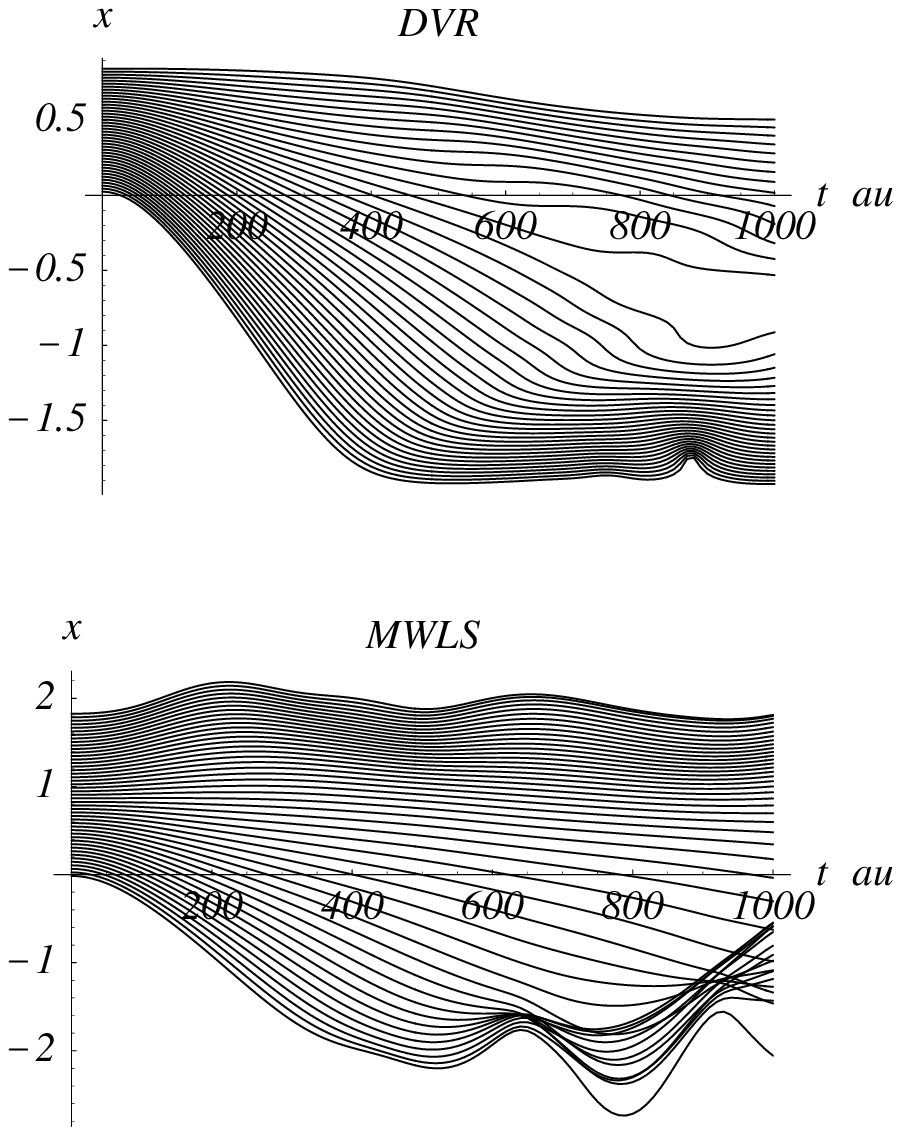}
\end{figure}

\begin{figure}
\caption{ Evolution of quantum density, $\rho$ (dots), quantum 
potential (short dashed lines),  $V(x)$ (solid line), and $V+Q (long 
dashed lines)$ 
for a proton in a double well potential at various times.  } 
\mathGraphic{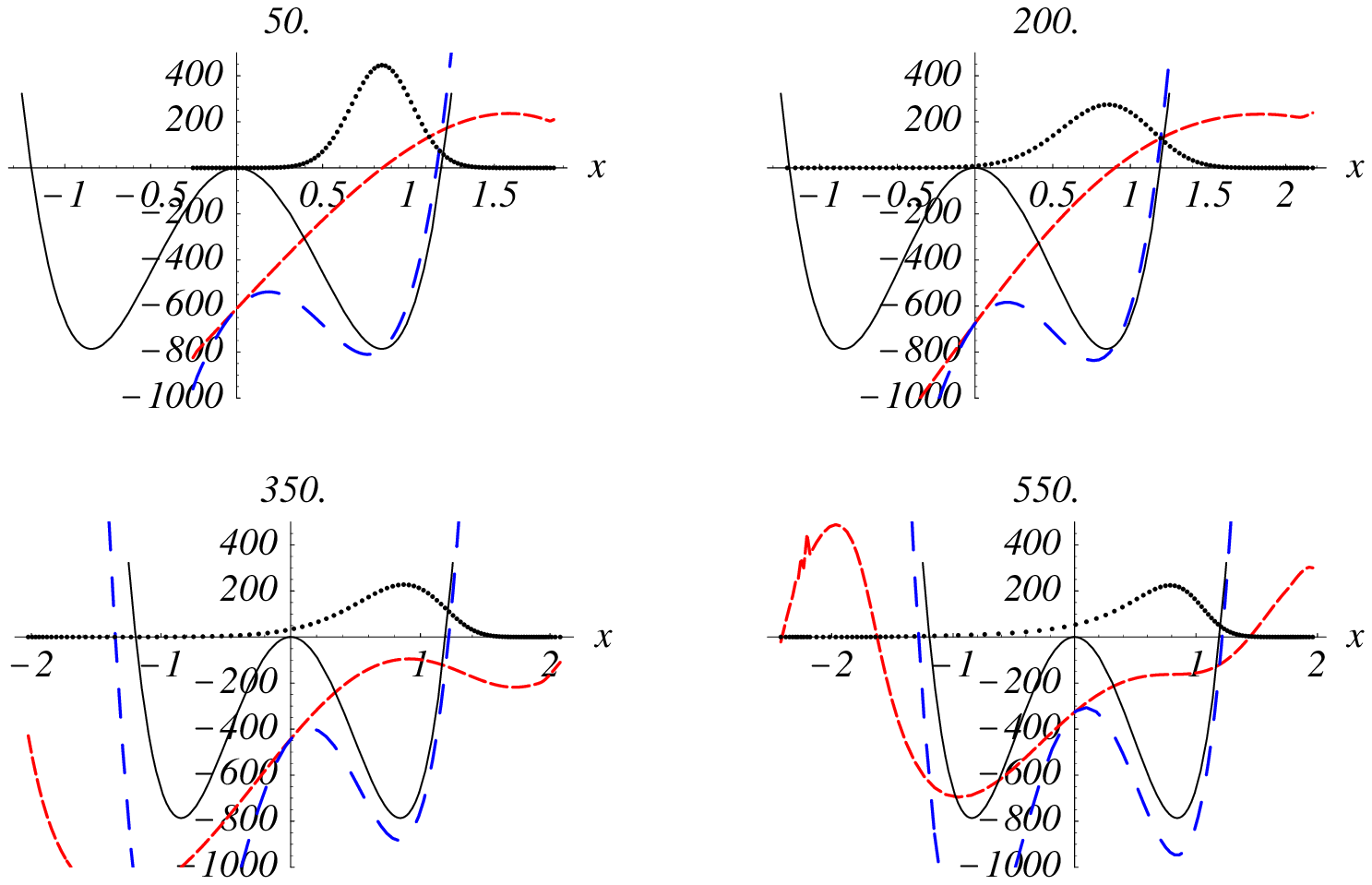}
\end{figure}

\begin{figure}
\caption{Effective barrier relative to bottom of well encountered by 
quantum trajectories in the double well problem in crossing from right 
to left hand well.}
\mathGraphic{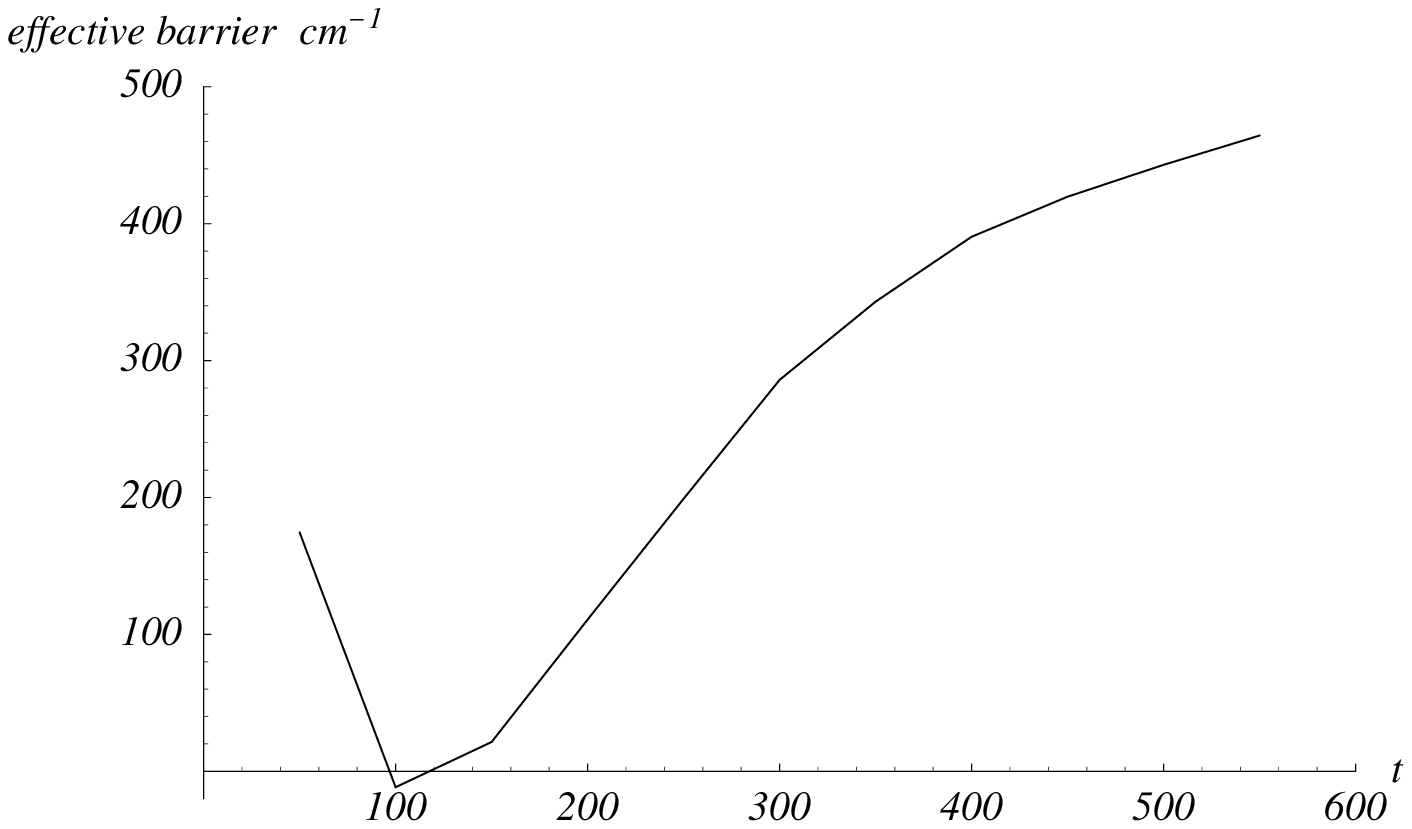}
\end{figure}

\begin{figure}
\caption{Evolution of the quantum density, $\rho$, as computed via MWLS based methods.}
\mathGraphic{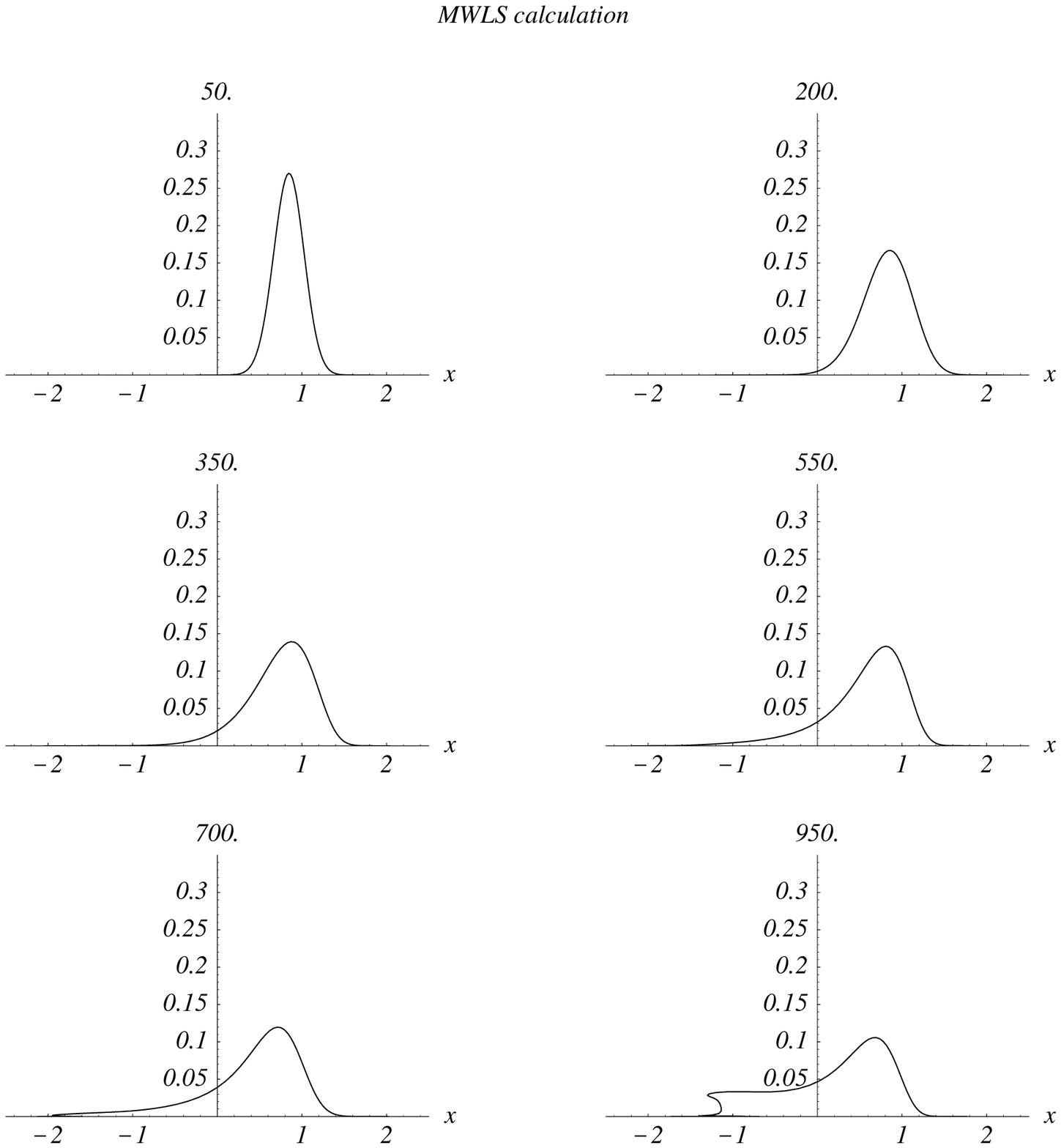}
\end{figure}

\begin{figure}
\caption{ Evolution of the quantum density, $\rho$, as computed via DVR 
based methods.  Note in particular the interference structure which 
develops in the DVR calculation which does not appear in the MWLS 
trajectory calculations shown in Figure 5.}

\mathGraphic{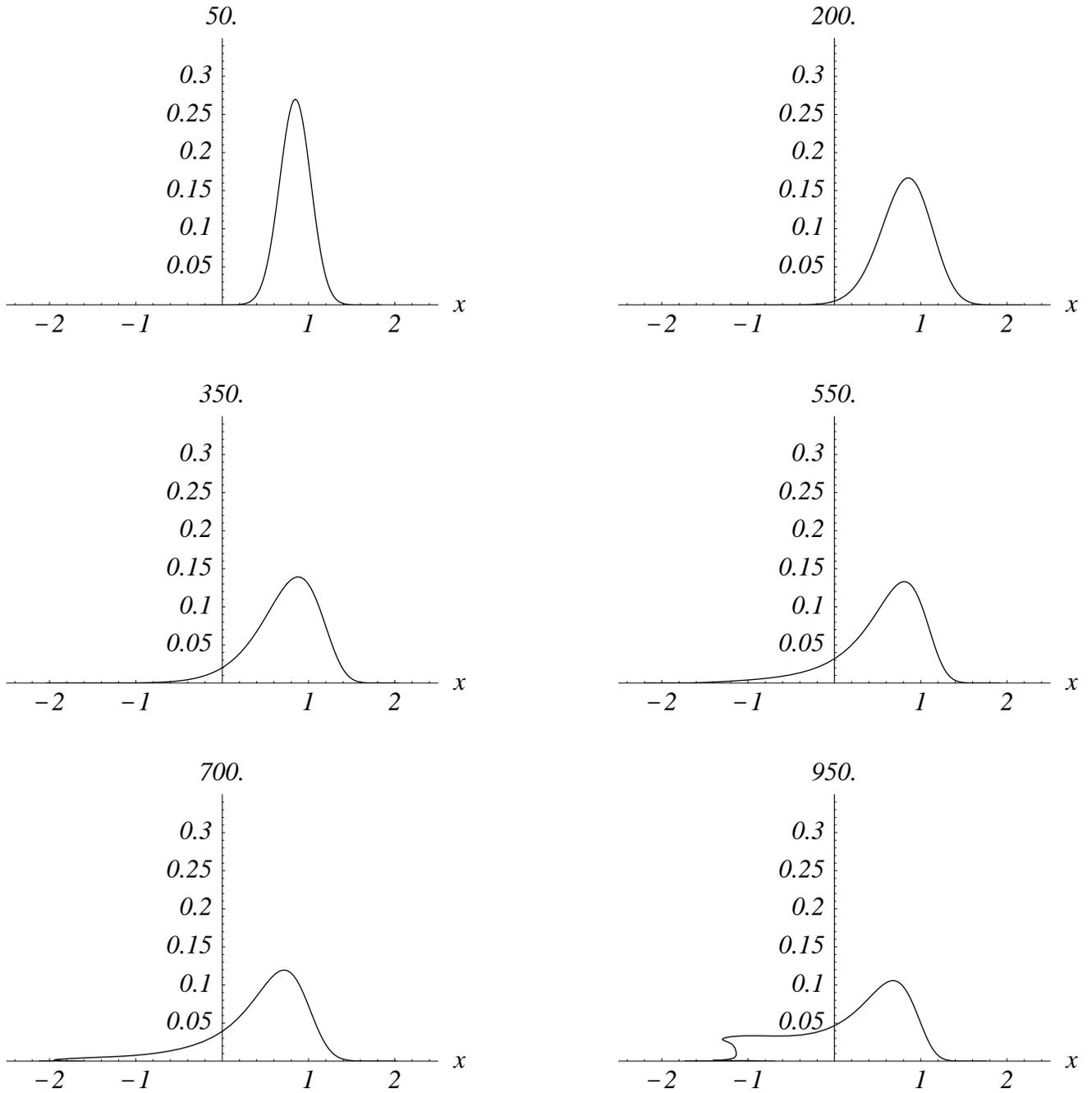}
\end{figure}

\begin{figure}
\caption{Comparison of 4 trajectories computed via DVR (dashed) versus MWLS (solid) methods.}
\mathGraphic{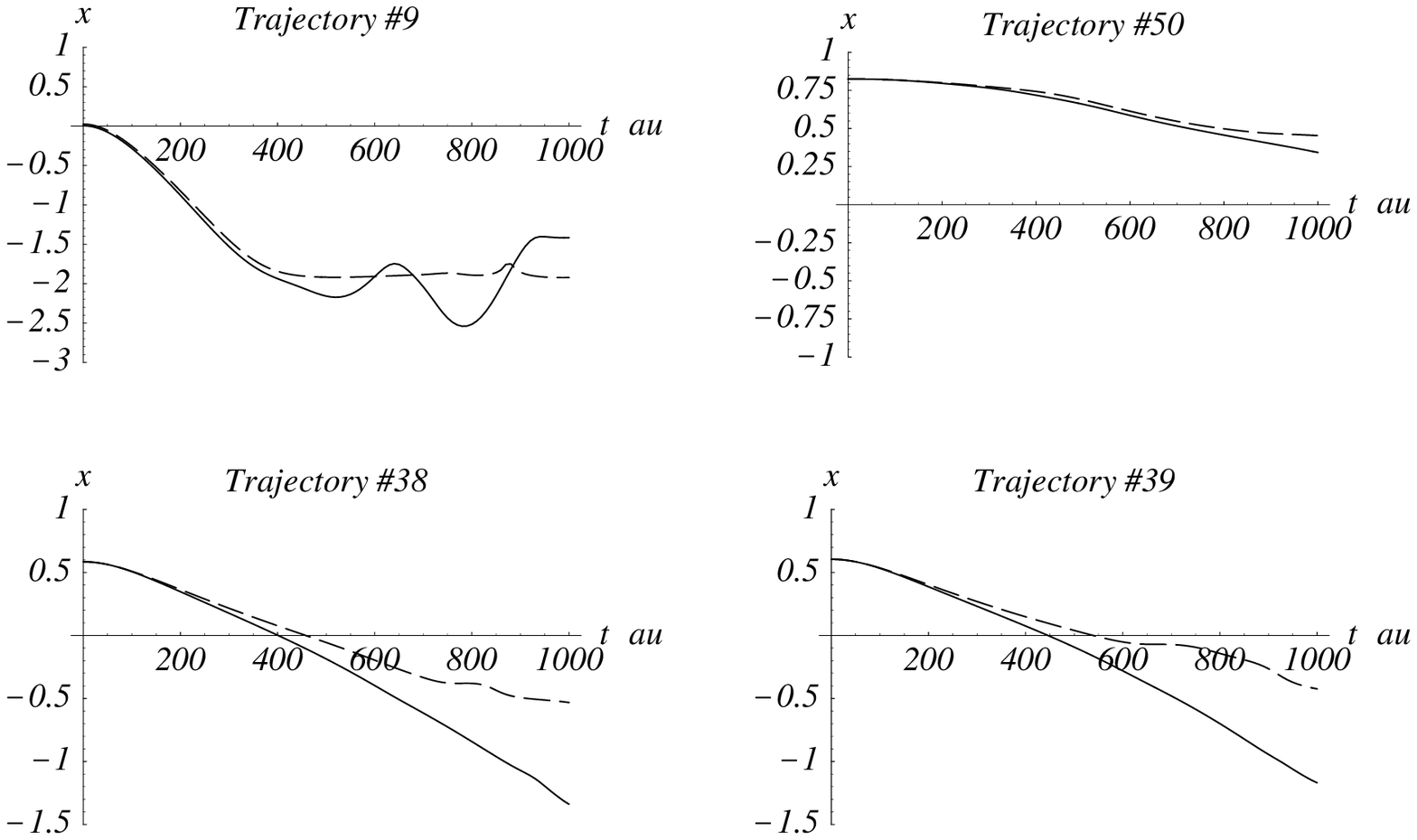}
\end{figure}


\begin{thebibliography}{10}
\bibitem{ref24}
Similar thoughts along these lines regarding the "node" problem are presented in Ref. 16.
\end{thebibliography}
\end{document}